# The effect of hole-doping on the magnetic properties of $Nd_{1-x}TiO_3$

Athena S. Sefat, John E. Greedan

Brockhouse Institute for Materials Research and Department of Chemistry, McMaster University, Hamilton, Ontario, Canada

Lachlan Cranswick

Canadian Neutron Beam Centre, Chalk River, Ontario, Canada

**Abstract**

The magnetic properties of the hole($x$)-doped Mott-Hubbard, antiferromagnetic insulator, $NdTiO_3$, have been investigated in $Nd_{1-x}TiO_3$. For the materials in $0.010(6) \leq x \leq 0.112(4)$, correlation between the structural and magnetic properties is discussed with respect to the Ti – O – Ti angles and the Ti – O bond distances. For $0.010(6) \leq x \leq 0.071(10)$, long-range magnetic order is observed through low temperature neutron diffraction and heat capacity. Néel ordering temperatures of 88.2 K and 61.9 K are determined, respectively, for $x = 0.019(6)$ and $x = 0.064(10)$. From high resolution neutron diffraction, the *Pnma* magnetic structure is assigned, unambiguously, as $G_xC_y$ on $Ti^{3+}$ and $C_y^2$ on $Nd^{3+}$. The $Ti^{3+}$ ordered moment decreases gradually from 0.45(8) $\mu_B$ for $x = 0.019(6)$ to 0.31(7) $\mu_B$ for $x = 0.071(10)$, but vanishes abruptly for higher $x$. Antiferromagnetic short-range order exists for $0.074(9) \leq x < 0.098(10)$ as determined by neutron diffraction and the field-dependence of the susceptibility. This is the first observation of short-range order for any hole-doped titanate system. For 29(4)% hole-doping in $Nd_{1-x}TiO_3$, $x \geq 0.098(10)$, all magnetic ordering is destroyed.

# I. INTRODUCTION

The renewed interest in lanthanide titanate perovskites has been inspired by analogy with the cuprate superconductors with a nominal $^2D$ ($3d^9$) $Cu^{2+}$ free ion ground state. Most of the pure, parent materials of the high-$T_c$ superconductors are antiferromagnetic (AF) insulators and electronic conduction is established by doping with, for example, $Sr^{2+}$ in $La_{2-x}Sr_xCuO_4$ (LASCO), $O^{2-}$ in $YBa_2Cu_3O_{6+y}$ (YBCO) [hole doping] or $Ce^{4+}$ in $Nd_{2-x}Ce_xCuO_4$ [electron doping]. For each of these systems, doping of the parent charge-transfer insulator results in a rapid decrease of the Néel temperature and ultimate quenching of the AF order, followed by a crossover to the superconducting state through a spin-glass phase [1]. The magnetic properties of transition metal oxides, like the perovskite titanates, $R$TiO$_3$ with $R$ = La – Gd, have been under study for many years. For early transition metal oxides, like $R$TiO$_3$, the $t_{2g}$ level is partially filled and the hybridization with oxygen ligands is very weak and the materials are generally known as Mott-Hubbard insulators. For Mott-Hubbard insulators, the charge excitation gap is formed between the lower Mott-Hubbard $d$-band with single occupancy of the atomic sites and the upper Mott-Hubbard band. For the late transition-metal oxides, like La$_2$CuO$_4$, the $t_{2g}$ level is completely filled and the Fermi level lies within the $e_g$ band in the absence of correlation, $U$. As a result, hybridization with the ligand is stronger and the oxygen $2p$ level is closer to the partially filled $e_g$ band. Upon the introduction of correlation, the $e_g$ band is split into a filled lower Hubbard band and an empty upper Hubbard band and the lowest lying charge excitation is from the ligand $p$-band to the upper-Hubbard $d$-band. These distinctions have been described in detail elsewhere [2, 3]. The members of the $R$TiO$_3$ series are one electron, $3d^1$, analogs to the superconducting cuprates with a single hole in the $3d$ shell. For $R$TiO$_3$, the mismatch of the $R$ ionic radii with the size of the available site in the structure causes tilting of the TiO$_6$ octahedra and induces the GdFeO$_3$-type lattice distortion (Figure 1). The decrease in $R^{3+}$ ionic radii across the lanthanide series leads to increased internal stress in the crystal structure that is accommodated by a decrease of Ti – O – Ti bond angle which results in a reduction of the width of the lower-Hubbard band, $W$ [4]. The parent LaTiO$_3$ and YTiO$_3$ compounds have Mott-Hubbard gaps ($E_g$) of 0.2 eV and 1 eV, respectively [5]. An intra-atomic Coulomb repulsion energy ($U$) of 4.0 eV has been reported for LaTiO$_3$ and La$_{0.90}$Sr$_{0.10}$TiO$_3$ using resonant soft-x-ray emission (SXES) in combination with photoemission and inverse-photoemission spectroscopy (PES/IPES) [6]. $U$ can be viewed as a constant throughout the $R$TiO$_3$ series [4]. In our companion study of optical conductivity on Nd$_{1-x}$TiO$_3$, we have found that NdTiO$_3$ is indeed a Mott-Hubbard insulator in which the lowest energy transition at ~ 0.8 eV is clearly associated with the Mott-Hubbard gap while the charge-transfer excitation occurs at 4 eV [7]. We have estimated the full width of the lower Mott-Hubbard band as 3.2 eV for Nd$_{0.981(6)}$TiO$_3$ [7], which is slightly larger than the calculated $W$ = 2.27 eV reported for NdTiO$_3$ using the tight-binding model [4].

The early La, Ce, Pr, Nd and Sm lanthanide titanates are canted-antiferromagnets due to the antisymmetric Dzyaloshinsky-Moriya interaction and order below ~ 146 K, 116 K, 120 K, 95 K and 52 K, respectively [8 - 11]. In these compounds, the Néel temperature decreases as the $R$ ionic radius decreases across the lanthanide series and subsequently leads to ferromagnetic (FM) ordering among the Ti spins, below 30 K, in GdTiO$_3$ [12]. For the lanthanide titanates with larger lattice distortions, the ferromagnetic ground state is relatively well understood [13]. For LaTiO$_3$, however, the small ordered Ti$^{3+}$ moment of ~ 0.5 $\mu_B$ [9, 14], compared to the expected moment of ~ 1 $\mu_B$ [15] has long been puzzling and controversial. Some theoretical and experimental studies of perovskite titanates have been reported from the view point of the role of orbital degrees of freedom in determining the magnetic ground state [16]. For example, orbital ordering was observed in YTiO$_3$, from resonant x-ray scattering. Through a recent report of first-principles calculations (LDA-DMFT) [17], it has been found that the lifting of $t_{2g}$-degeneracy in YTiO$_3$ arises primarily via a Jahn Teller distortion [14, 18] while in LaTiO$_3$, the lifting of $t_{2g}$-degenercy is ascribed to the crystal fields from the La-cations. In LaTiO$_3$, the GdFeO$_3$-type distortion tilts and rotates the corner-sharing octahedra as illustrated in Figure 1. It has been shown that this distortion is partly driven by the covalency between occupied oxygen $p$ states and empty La-cation $d$ states, which pull each O1 (O2) closer to one (two) of its four nearest La-neighbors [19]. As a result, each La-cation has 4 of its 12 near oxygens pulled closer in. In addition, the La-cubeoctahedron becomes distorted so that one diagonal becomes the shortest.

There have been several studies of the effect of hole doping on the magnetic properties of AF rare earth titanites for example $La_{1-x}Sr_xTiO_3$ [11], $La_{1-x}TiO_3$ [20, 21], $LaTiO_{3+\delta}$ [22], $Nd_{1-x}A_xTiO_3$ ($A$ = Ca, Sr, Ba) [23], $R_{1-x}Ca_xTiO_3$ ($R$ = La, Pr, Nd and Sm) [4] and $Nd_{1-x}TiO_3$ [24, 25]. With sufficient doping, these insulating $R$TiO$_3$ materials are driven metallic and the long range AF order is quenched. In most cases the Néel temperature, $T_N$, has been monitored as a function of doping level. In general, two indirect methods have been used to determine $T_N$: the observation of the zfc/fc divergence in the d.c. susceptibility and the disappearance of the remanent magnetization in zero applied field upon heating. In the present work, the accuracy of these methods is evaluated by comparison with $T_N$ values determined by heat capacity measurements. In two cases, $LaTiO_{3+\delta}$ and $Nd_{1-x}TiO_3$, however, the ordered moment on $Ti^{3+}$ has been measured and $T_N$ followed by neutron diffraction [22, 25]. In all previous studies the hole concentration grid was fairly coarse, with only a few samples, two or three, spanning the Mott transition. In comparison with such published reports, the present study for $Nd_{1-x}TiO_3$ is more detailed, involving more than ten samples, well characterized with respect to composition, over the range $0.010(6) \leq x \leq 0.112(4)$ which includes the Mott-Anderson transition, as we reported recently [26]. The Néel temperature is monitored using heat capacity, d.c. susceptibility, zfc/fc divergence and remanent magnetization methods and the results are comparatively evaluated. As well, the ordered moments on both $Ti^{3+}$ and $Nd^{3+}$ are determined, using high resolution neutron diffraction, also as a function of hole($x$)-doping levels. These results will be combined with our studies of optical and electrical transport properties on the same materials [7, 26]. It is important to determine if the antiferromagnetic ordered state persists into the metallic regime as reported in $Nd_{1-x}Ca_xTiO_3$ system [4].

## II. EXPERIMENTS

The polycrystalline $Nd_{1-x}TiO_3$ samples with $0.010(6) \leq x \leq 0.112(4)$ were prepared by mixing, grinding and pelletizing stoichiometric amounts of $Ti_2O_3$ (Cerac, 99.9%), pre-dried $Nd_2O_3$ (Research Chemicals 99.99%) and $TiO_2$ (Fisher Scientific 99.97%). Each sample was sealed in a molybdenum crucible under purified argon gas. The preparation conditions involved several firing steps, at ~1400 °C for ~12 hours in an *rf* induction furnace. The single crystal composition of $x = 0.064(10)$ was grown using the Bridgeman method with uniform melting of the raw materials in a molybdenum crucible and slow cooling in a temperature-gradient induction furnace. The single crystal growth for $x = 0.019(6)$ was carried out by a floating zone (FZ) technique using a double ellipsoid image furnace (NEC SCI-MDH-11020). For this growth run, the translational velocity of the seed rod was launched at ~ 25 mm/hour and the experiment was performed in ~ 3.5 atm flow of 5% $H_2$/argon gas mixtures.

Phase purity of the samples was initially monitored by x-ray powder diffraction using a Guinier-Hägg camera with monochromated Cu $K\alpha_1$ radiation and silicon standard. As well, very accurate and precise unit cell constants were obtained from the Guinier-Hägg data. Neutron activation analyses of the sintered samples were used to fix the vacancy levels and the weight gains were used to verify the $Nd^{3+}$ present per unit formula (see Section III). The neutron activation analyses were performed at the McMaster Nuclear Reactor and the thermogravimetric analyses were done using 409-Netzch or PC-Luxx Simultaneous Thermal Analyzer.

DC Magnetization of each sample was measured as a function of temperature and magnetic field using a Quantum Design Magnetic Property Measurement System, in the temperature range of 2 - 300 K and 0 - 2 T. In a typical susceptibility experiment, the sample was cooled to 2 K in zero-field (zfc) and the data were then collected from 2 K to 300 K with an applied field. The sample was subsequently cooled in the applied field, or field-cooled (fc), and the measurements were repeated from 2 K to 300 K. For testing the notion of small ferromagnetic ordering in $Nd_{1-x}TiO_3$ samples, zero-field measurements were conducted on numerous compositions. The samples were cooled from room temperature to 2 K in a magnetic field of 5 T to saturate a possible ferromagnetic moment. The field was then turned off at 2 K and magnetization measurements were recorded as the samples were warmed in zero field to room temperature.

Neutron Powder Diffraction data were collected on the C2 diffractometer operated by the Neutron Program for Materials Research of the National Research Council of Canada at Chalk River Laboratories of AECL Ltd. The samples of ~ 1- 2 g were placed in helium-filled vanadium cans that were sealed with an indium gasket. The chemical structural refinements for several samples with $0.019(6) \leq x \leq 0.098(10)$ were obtained from neutron powder data ($\lambda \cong 1.32$ Å) collected at room temperature, over the range $10° \leq 2\theta \leq 115°$. For the magnetic structure determinations, data were collected at ~4 K using ($\lambda \cong 2.37$ Å.) neutrons, for $0.010(6) \leq x \leq 0.112(4)$ samples. Refinements of both the chemical and magnetic structures were carried out using the Fullprof suite of programs, WINPLOTR [27]. The Bragg peak shapes were modeled using the Pseudo-Voigt (PV) function convoluted with an axial divergence asymmetry function [28].

The heat capacity data were collected on single crystal samples of $x = 0.019(6)$ and $x = 0.064(10)$. The calorimeter probe of the Oxford Instruments Maglab system was used for the measurements. The sample sizes were ~10 mg with dimensions of approximately 2 mm x 2 mm and <1 mm thickness. The sample bar was attached to the probe using a weighed amount of Wakefield grease. The relaxation method, at the applied fields of 0 T and 6 T was used for the measurements over a temperature range of 5 to ~ 110 K.

### III. COMPOSITIONAL ANALYSES

It is of course of critical importance to measure the vacancy concentration, $x$, in $Nd_{1-x}TiO_3$. For this purpose neutron activation (NAA) and thermal gravimetric (TGA) analyses were used. In the NAA technique the quantitative gamma emissions of the isotopes $^{149}Nd$, $^{151}Nd$ and $^{51}Ti$ from irradiated, polycrystalline $Nd_{1-x}TiO_3$ samples were compared to those obtained from reference samples with known Nd/Ti ratios. Three independent measurements of each sample were averaged for the references and unknowns. Thermogravimetic measurements were performed on the same batch of samples by heating in air to 1000°C. Having a measure of the Nd/Ti value from NAA measurements, the oxidative weight gain of each sample to $Nd_2Ti_2O_7$ and $TiO_2$ was monitored. Representing the system as $Nd_{1-x}TiO_{3+y}$, the effective titanium valence ($v$) and the hole filling ($n$) of the 3$d$ band can be given by the relations $v = 3 + 3x + 2y$ and $n = 1 - 3x - 2y$, respectively. In other words, off-stoichiometry on oxygen or neodymium can nominally introduce "holes" with concentration $\delta = 3x + 2y$ in the filled lower Mott Hubbard 3$d$ titanium band in $NdTiO_3$. In a first approximation one can set $y = 0$. On this basis, the theoretical and the observed TGA percent weight gains are listed Table 1 (columns $d$ and $e$). The discrepancies which exist can be accounted for by assuming small deviations from the nominal Nd/Ti ratio, column $f$ of Table 1, which lie within the uncertainty limits of the NAA technique. Thus, evidence from the NAA and TGA data indicates that $y \cong 0$ in the $Nd_{1-x}TiO_{3+y}$ series. This cation vacancy model makes sense on structural grounds, since it is hard to visualize where the large $O^{2-}$ ion could be forced into the close-packed perovskite lattice and difficult to understand how such large interstitial ions could cause the unit cell to shrink with increasing level of oxidation. Also, the perovskite structure is maintained down to $x = 0.33$ ($Nd_{2/3}TiO_3$) and solid solutions can be successfully synthesized keeping the Ti – O network intact. Still, some recent reports have ignored the $R$-vacancy model in favor of the excess oxygen model, e.g. $RTiO_{3+y}$, but without any elemental analysis [29]. Because NAA analyses could not be done on all $Nd_{1-x}TiO_3$ samples, the correlation between the Nd/Ti ratio from NAA and the unit cell volume determined from Guinier powder diffraction, Figure 2, was used to assign additional doping levels.

## IV. STRUCTURAL RESULTS

Two structural types have been identified for $Nd_{1-x}TiO_3$ neodymium-deficient system: *Pnma* for 0 ≤ *x* ≤ 0.25 [20] and *Cmmm* for *x* = 0.30 [30]. In this study we deal only with compositions having the *Pnma* form (Figure 1) and 0.019(6) ≤ *x* ≤ 0.098(10). The Rietveld refinement for *x* = 0.019(6) is displayed in Figure 3, as an example. The agreement factors and refined atomic positions can be found in Tables 2 and 3, respectively. For $Nd_{1-x}TiO_3$, the trends in Ti – O bond distances and Ti – O – Ti bond angles with *x* are presented in Figure 4.

For the $Nd_{1-x}TiO_3$ solid solution, one anticipates a strong correlation between the magnitude of the distortion of the $TiO_6$ octahedra, the Ti – O – Ti angles and the physical properties. For $Nd_{0.981(6)}TiO_3$ with *x* = 0.019(6), an average Ti – O – Ti bond angle of 149.0(4)° was found (Figure 4). With increasing *x*, the expected decrease in the average Ti – O bond distances (as the smaller $Ti^{4+}$ substitutes for $Ti^{3+}$) is coupled with an almost linear increase in the average Ti – O – Ti bond angle. Because the titanium $3d^1$ electron transfer in $Nd_{1-x}TiO_3$ is governed by the superexchange process mediated by the O 2*p* states, these trends should be taken into account when the magnetic properties are considered. For example, one expects that the strength of the superexchange interactions coupling the $Ti^{3+}$ moments will increase as the Ti – O – Ti angles increase. In the undoped *R*TiO_3 materials, $T_N$ increases as the Ti – O – Ti angle increases [31].

## V. MAGNETIC PROPERTIES

A major goal of this work is to determine the change in Néel ordering temperature as a function of the hole-doping level in $Nd_{1-x}TiO_3$. Emphasis has been placed on techniques which are considered to be the most reliable namely, the thermal heat capacity, the so-called Fisher heat capacity and neutron diffraction. Given the small ordered moments, the latter technique is not practical for these materials. The results are compared with those obtained by alternative methods which are often used, being the zfc/fc divergence in the d.c. susceptibility and the disappearance upon heating of the remanent magnetization.

## V. I. Monitoring $T_N$ versus $x$

### V.I. I. Heat capacity and Fisher's heat capacity

For the spin ½ system $Nd_{1-x}TiO_3$, the magnetic contribution to the heat capacity due to the canted-antiferromagnetic ordering of $Ti^{3+}$ moments, at the Néel temperature, is expected to be small. The $Nd^{3+}$ moments were previously reported to order independently only at ~1 K [24]. Neodymium magnetic ordering has also been observed in $NdMO_3$ ($M$ = Co, Fe, Cr, Ni) in the 1 K range [32]. Here, we focus on finding the Néel magnetic transition due to titanium ordering, as a function of $x$, and compare the temperature for this anomaly in the heat capacity to the features manifested in the susceptibility and magnetization data.

The specific heat data were collected on single crystal samples of $Nd_{0.981(6)}TiO_3$ ($x$ = 0.019), $Nd_{0.936(10)}TiO_3$ ($x$ = 0.064) and $Nd_{0.89}TiO_3$ ($x$ = 0.11). For $x$ = 0.019(6) and $x$ = 0.064(10), continuous lambda-like features are evident in Figure 5 with peaks at 88.3 K and 61.9 K, respectively. Such anomalies are indicative of transitions to long-range order (LRO) and are assigned as Néel temperatures. The paramagnetic $x$ = 0.11 sample (Figure 5) was taken as the lattice contribution to heat capacity and was subtracted from the total heat capacity signal of the $x$ = 0.019(6) and $x$ = 0.064(10) samples. The magnetic heat capacities of these samples are illustrated in Figure 6; Figure 6b also depicts measurements in an applied field of 6 T for $x$ = 0.064(10). Note that there appears to be no change in the magnetic heat capacity which is not surprising given the high $T_N$ value for this sample. As evident from Figure 6, the entropy loss is spread over a wide temperature range. The entropy values calculated over the temperature ranges are summarized in Table 4 for the two samples. The entropy release due to Néel magnetic ordering for the $x$ = 0.019(6) and $x$ = 0.064(10) samples are ~ 34% and 25%, respectively, compared to the ideal 5.7628 J/(mol K) for a spin ½ system and the appropriate concentration of $Ti^{3+}$ in the two samples.

Fisher has shown that $d(\chi T)/dT$ is a good approximation to the magnetic component of the thermal heat capacity [33]. The $T_N$'s from the real heat capacity results are compared to the Fisher heat capacities obtained from zfc data at various applied fields in Figure 7 and the agreement is excellent. Fisher heat capacities were determined for other compositions and the results are included in Table 5.

Finally, in addition to the peaks at 88.3 K and 61.9 K for $x = 0.019(6)$ and $x = 0.064(10)$ samples, respectively, there seems to be a weak but sharp anomaly at ~ 25 to 30 K, common to both samples. This feature may be due to spin-reorientation of the coupled $Ti^{3+}/Nd^{3+}$ moments, similar to the isostructural $NdCrO_3$ in which a spin reorientation transition occurs for Cr spins at 34.2(5) K [34]. This possibility will be investigated using neutron diffraction, see section V.I.IV).

### V.I.II. Magnetic susceptibility: the zfc/fc Divergence

In the lanthanide titanates, $RTiO_3$, the canted-antiferromagnetic ordering gives rise to a weak ferromagnetic component [31]. The temperature at which the zfc and fc susceptibilities diverge ($T_D$) is often identified with $T_N$ in canted AF systems as in some previous reports on titanates, where the zfc/fc susceptibilities were measured at one particular applied field [25, 35]. However, our detailed exploration of this approach shows that this is not a reliable method of determining the long-range Néel ordering. In the present study, $\chi(T)$ has been measured at numerous applied fields for a wide range of $x$-doping levels. As evident from Figure 8, for example, the divergence temperature can be field-dependent. Even for the lightly doped $x = 0.019(6)$ sample in Figure 8a, where there is negligible effect of field on the divergence temperature, note that the observed $T_D$ at ~ 93 K is higher than the position of the lambda peak at 88.3 K in the heat capacity, Figure 6a. Interestingly, $T_D$ appears to coincide with the onset of the lambda peak, suggesting that it is sensitive to the formation of short-range spin clusters in the pre-critical regime, rather than to $T_N$. As the doping levels increase, the effect of applied field on $T_D$ becomes severe. For the range $0.010(6) \leq x \leq 0.071(10)$, the divergence temperature is nearly constant in various applied fields; however for $0.074(9) \leq x \leq 0.080(10)$, $T_D$ becomes a strong function of the applied field (Fig. 8b, c). This indicates that magnetic-short range ordering becomes more important as the doping level increases.

In the doping range $0.079(2) \leq x \leq 0.089(10)$, the zfc/fc divergences are very weak (Figure 9a, b). Finally, for $x = 0.098(10)$, the divergence vanishes and simple paramagnetic behavior is seen to 2 K (Figure 9c).

As canted AF materials should show hysteresis effects, the magnetic moment versus applied field was measured for $Nd_{0.981(6)}TiO_3$ ($x = 0.019$) at 2 K (Figure 8a, inset). Saturation is not reached at 1 T, however a weak hysteresis is evident. The remanent magnetization due to field of 1 T was found to be $\sim 4 \times 10^{-4}\ \mu_B$.

### V.I.III. Remanent magnetization

In another effort to locate the Néel ordering temperature ($T_N$) for the $Nd_{1-x}TiO_3$ system, the decay of the field induced magnetization was monitored as a function of temperature. The properties of a canted antiferromagnetic sample resemble those of a ferromagnet but with a much reduced spontaneous moment. In such materials, the magnetization strongly depends on magnetic history and it is possible to observe remanent moments in zero field. The magnetization results are shown in Figure 10 for numerous samples with $0.010(10) \leq x \leq 0.079(2)$. The titanium FM moment in $Nd_{0.990(6)}TiO_3$, with $x = 0.010(6)$, is $1.2 \times 10^{-2}\ \mu_B$; this value is comparable to the reported $2 \times 10^{-2}\ \mu_B$ for $Nd_{0.9}Ca_{0.1}TiO_3$, which was found at 5 K by cooling in the applied field of 3.5 T [36].

The temperature at which the remanent moment disappears for each sample (Figure 10) has been correlated with the $T_N$ in the literature, as for example in the $Nd_{1-x}Ca_xTiO_3$ system for $0 \leq x \leq 0.2$. In comparison with the $T_N$ values obtained from both the real heat capacity and Fisher's heat capacity in our study (Figure 7 and Table 5), there is reasonable agreement in some cases but for others, higher values are often found by the remanent moment analysis. For example for $x = 0.019(6)$, the remanent moment dies off at 87 K (see Figure 10), and the real and Fisher heat capacities show $T_N$ values of 88 K. On the other hand, for $x = 0.064(10)$, the remanent moment value of 66 K is higher than the heat capacity values which are both near 62 K. It is clear that below $x = 0.071(10)$, the apparent ordering temperature drops sharply, for example for $x = 0.079(2)$, there is evidence of a small remanent moment below ~35 K. Of course one expects a decrease in $T_N$ with hole doping due in part to the dilution of $Ti^{3+}$ in the system. The data of Figure 4 indicate, however, that structural changes which accompany the hole doping are likely to mitigate the effects of dilution to some extent. For example the average Ti – O – Ti angle, which is involved in the superexchange spin transfers, actually increases with increasing $x$. From the strong increase of $T_N$ with increasing Ti – O – Ti angle in the pure $R$TiO$_3$ phases, one expects the strength of the $Ti^{3+}$ superexchange coupling to increase with doping level in the Nd$_{1-x}$TiO$_3$ system.

### V. I.IV. Thermal neutron diffraction: The Magnetic Structure of Nd$_{0.981(6)}$TiO$_3$

As mentioned, the $R$TiO$_3$ perovskite compounds exhibit a remarkable cross over from from Ti - Ti AF coupling in the lighter rare earths ($R$ = La – Sm) to FM Ti - Ti coupling for the $R$ = Gd - Lu, Y compounds. In the older literature, the magnetic structures have been discussed in the non-standard space group *Pbnm* while in this work, the standard setting, *Pnma* is used. The magnetic structure of LaTiO$_3$ was reported to be composed of a $G_xF_z$ spin configuration on the titanium, *Pbnm* setting, ($G_zF_y$ in *Pnma*) or a $G_zF_x$ ($G_xF_z$ in *Pnma*); it has been difficult to distinguish between these due to either low resolution in powder studies or twining in single crystal studies [9, 22]. The $G_xF_z$ notation follows Bertaut [37] and indicates that a ferromagnetic component along the z-direction can only be coupled with a $G$-type mode along the x-axis. The magnetic structures of $R$TiO$_3$ with $R$ = Ce, Pr and Nd have been ascribed to a $G_zF_x$ spin configuration ($G_yF_z$ in *Pnma*) on $Ti^{3+}$ sublattice and $F_xC_y$ spin configuration ($F_zC_x$ in *Pnma*) on the rare-earth sublattices [9, 14, 25]; the weak ferromagnetic components in these samples were presumed to arise from the canting of the antiferromagnetic moments on $Ti^{3+}$ as the F component is too weak to be detected by neutron diffraction.

Here, the magnetic structure of Nd$_{0.981(6)}$TiO$_3$, with $x = 0.019(6)$, was refined using high resolution powder neutron data and the program Fullprof [27]. The magnetic structure described previously for NdTiO$_3$ was taken as the initial model [25] and then several other models which placed the moments along different axes, for example Ti-$G_y$, Nd-$C_x$, and Ti-$G_x$, Nd-$C_y$ were tested. The refined profile for $x = 0.019(6)$ in *Pnma* is displayed in Figure 11 wherein the strong magnetic reflections are identified with arrows. Note that two resolved doublets are present. That at lower angles is comprised of the (100) & (001) reflections which are due mainly to the AF type-$C$ $Nd^{3+}$ structure and at higher angles, the (110) & (011) reflections, which arise from the AF type-$G_x$ $Ti^{3+}$ structure. The refinements showed, unequivocally, that the moment confirgurations are $G_x$(Ti) and $C_y$(Nd) which is depicted in Figure 12. To our knowledge, this is the first unambiguous determination of the magnetic structure from powder data for any antiferromagnetic $R$TiO$_3$ material. Note that this result is different from that reported previously for NdTiO$_3$ [25]. Interestingly, for these spin configurations of $Ti^{3+}$ and $Nd^{3+}$ at 4 K, a ferromagnetic component is symmetry forbidden [37]. This is surprising given that the anti-symmetric Dzialoshinski-Moriya interaction [38] is allowed in these materials and that a ferromagnetic component is apparently seen near $T_N$ in the form of a zfc/fc divergence.

Close inspection of the (110) and (011) doublet in the neutron data for $x = 0.019(6)$ over the temperature range of 10 to 40 K showed no evidence of a spin re-orientation on the $Ti^{3+}$ site. As a result, the origin of the low temperature heat capacity anomaly at ~ 25 - 30 K (Figure 6) is still unknown. The thermal development of the $Ti^{3+}$ moment was followed over the range 4 K ≤ T ≤ 110 K, Figure 13. The results are consistent with $T_N$ of 88.3 K in agreement with the heat capacity data and as well the Ti-$G_x$, Nd-$C_y$ model. Note that the temperature dependence of the $Nd^{3+}$ moment shows that it is induced by strong $Ti^{3+}$- $Nd^{3+}$ coupling.

## V. II. Monitoring the Ti$^{3+}$ ordered moment vs *x*

The evolution of the long-range ordered moment on both $Ti^{3+}$ and $Nd^{3+}$ in the Nd$_{1-x}$TiO$_3$ system is obtained here through neutron diffraction measurements of numerous vacancy-doped compositions at 4 K. Magnetic Bragg peaks were detected for samples with $x \leq 0.071(10)$. The conditions and refinement results for the combined chemical and magnetic structures for $Nd_{1-x}TiO_3$ samples are summarized in Table 6. The 4 K neutron diffraction experiments found long-range AF order for $0.019(6) \leq x \leq 0.071(10)$ with a finite moment on $Ti^{3+}$. The refined magnetic moments per $Ti^{3+}$ and $Nd^{3+}$ for these $Nd_{1-x}TiO_3$ compositions are plotted in Figure 14 along with the values for the $x = 0.074(9)$ sample. As indicated previously, the $Nd^{3+}$ moment arises only through coupling with $Ti^{3+}$. Note that both moments show a rather weak dependence on the doping level to x = 0.071. The refined $Ti^{3+}$ moment for $x = 0.074(9)$ is very small with a large error and a large $R_{mag}$ value. Thus, while it is not clear from a direct measurement whether an ordered moment still exists on $Ti^{3+}$, the observation of a finite moment on $Nd^{3+}$ implies such for this composition. For the $x = 0.079(2)$ and $0.089(1)$ samples, the magnetic reflections associated with the $Nd^{3+}$ moments are clearly broadened, indicative of short-range AF ordering (Figure 15). Note that this implies that a short range ordered moment must still exist on the $Ti^{3+}$ sublattice, as, in the absence of a $Ti^{3+}$ moment, the $Nd^{3+}$ moments order only below 1 K [24]. This is corroborated by the observation of a zfc/fc divergence in the susceptibility, Figure 9a, b. For $x = 0.095(8)$, there is no evidence of magnetic order down to 4 K in agreement with the results of Figure 9c. This is the first report of a regime of short range magnetic order in a hole doped perovskite titanate system. Further investigations into the nature of the short range ordering are in progress.

The small $Ti^{3+}$ ordered moment of 0.45(7) $\mu_B$ in $Nd_{0.981(6)}TiO_3$ (Table 6) is similar to the reported values of 0.46(2) $\mu_B$ for $LaTiO_3$ at ~ 5 K [22, 9], 0.4 $\mu_B$ for $CeTiO_3$ and 0.25 $\mu_B$ for $PrTiO_3$ [9]. Our result, however, contradicts the ordered titanium moment of 0.99(5) $\mu_B$ reported previously for $NdTiO_3$ [25]. We believe that the low resolution of the data in this particular report may have contributed to the finding of a false minimum and to an overestimation of the moment. Subsequent refinement of the same data used in [28] with the model derived from our high resolution data yielded moments of 0.535(54) $\mu_B$ on $Ti^{3+}$ and 0.889(33) $\mu_B$ on $Nd^{3+}$ with $G_x$- and $C_y$-type configurations, respectively. The very weak dependence of the $Ti^{3+}$ moment on doping level correlates well with the structural results, Figure 4. Much recent work has shown that the value of the $Ti^{3+}$ moment is a strong function of the details of the distortion of the Ti – O coordination octahedron, larger moments are associated with larger distortions, see for example references [22, 39]. The results show that the Ti – O environment changes very little over the range $x \leq 0.071(10)$.

Finally, in a companion paper we show that the onset of true metallic behavior over the entire temperature range occurs at $x = 0.098$. Samples within the range of $0.074(10) \leq x \leq 0.089(1)$ show non-metallic behavior below ~ 150 K [26]. Thus, it is possible to conclude that for the $Nd_{1-x}TiO_3$ system, a long range ordered AF state does not extend into the metallic regime, in sharp contrast to the report of such a state in the very similar $Nd_{1-x}Ca_xTiO_3$ system [4].

## VI. CONCLUSIONS

The effect of hole doping, via the introduction of $Nd^{3+}$ vacancies, on the magnetic properties of the Mott-Hubbard AF insulator $NdTiO_3$ has been studied in detail using a variety of experimental probes. In the present study, materials with vacancy concentrations $0.010(6) \leq x \leq 0.112(5)$ have been synthesized and characterized with respect to composition and structure. The introduction of each $Nd^{3+}$ vacancy introduces three holes on the $Ti^{3+}$ sites. All samples within this compositional range crystallize in the $GdFeO_3$ distorted perovskite structure described by space group *Pnma* and changes in the Ti – O distances and Ti – O – Ti angles with hole doping have been determined in detail. This paper is focussed on the consequences of hole-doping for the magnetic properties of $Nd_{1-x}TiO_3$ and the major conclusions are as follows. Firstly, the doping-induced changes in both the Neél temperature and the ordered magnetic moments on the $Ti^{3+}$ and $Nd^{3+}$ sites were accurately monitored. With respect to the Neél temperature determination, the accuracy of four methods was compared: these are the heat capacity, the susceptibility derivative or Fisher heat capacity, the zfc/fc susceptibility and the temperature dependence of the zero-field remanent magnetizations. The last two have been widely used in previous work on related systems [9, 10, 14, 40, 41], but the present results show that these to be more sensitive to the onset of short range, pre-critical ordering and generally to overestimate $T_N$. Thus, the heat capacity and Fisher heat capacity are concluded to be the most reliable probes for this purpose. $T_N$ is found to decrease approximately linearly with doping levels from 100 K for $x \sim 0$ to 60 K for $x = 0.071(10)$ and the results are shown in Figure 16. By combining observations from both heat capacity and neutron diffraction it was shown that true long-range order occurs only for $0.019(6) \leq x \leq 0.071(10)$ and that $T_N$ is completely quenched for $x > 0.074(9)$. In a companion paper, we show from electrical transport studies, that the fully metallic state is not reached until $x = 0.098(10)$ [26]. Thus, this observation is in sharp conflict with the conclusions drawn from a previous study of the $Nd_{1-x}Ca_xTiO_3$ system for which it was claimed that a long range AF state persisted into the metallic regime [4]. Secondly, the use of high resolution neutron data results in the first truly unambiguous determination of the magnetic spin configurations at $Ti^{3+}(G_x)$ and $Nd^{3+}(C_y)$ (in $Pbnm$ as $G_yC_z$) of any titanate perovskite from powder diffraction. The small $Ti^{3+}$ ordered moment, $0.45(7)$ $\mu_B$, in the parent $Nd_{0.981(6)}TiO_3$ is comparable to the values of $\sim 0.4$ $\mu_B$ reported for $LaTiO_3$ [9, 22], 0.4 $\mu_B$ for $CeTiO_3$ and 0.25 $\mu_B$ for $PrTiO_3$ [9] but contradicts the ordered titanium moment of $0.99(5)$ $\mu_B$ reported previously for $NdTiO_3$ [25]. A plausible argument was presented to account for this discrepancy. Thirdly, the ordered $Ti^{3+}$ moment was found to be nearly independent of $x$ in range $0.019(6) \leq x \leq 0.071(10)$ and then to plummet to $< 0.1$ $\mu_B$ at $x = 0.074$. These results should be compared with the only other similar study of which we are aware, that for oxidized forms of $LaTiO_3$, formulated as $LaTiO_{3+\delta}$ [22]. In this work the ordered moment at the Ti site, uncorrected for dilution, was found to decrease with increasing $\delta$ with moment values of $0.46(2)$ $\mu_B$ for $\delta = 0.000(5)$, $0.32(2)$ $\mu_B$ for $\delta = 0.030(5)$ and $0.09(3)$ $\mu_B$ for $\delta = 0.07(1)$. Direct comparison with our results requires conversion of $\delta$ into a cation deficiency, $x$, which is the preferred oxidative mechanism for titanate perovskites and correction for the dilution effect of $Ti^{4+}$. Thus, the $\delta = 0.030$ $LaTiO_{3+\delta}$ sample becomes $x \sim 0.02$ with a $Ti^{3+}$ moment of $0.33$ $\mu_B/$ $Ti^{3+}$ ion, and $\delta = 0.07$ is $x \sim 0.05$ with $0.10$ $\mu_B/$ $Ti^{3+}$ ion. These values are numerically close to those reported here for $Nd_{1-x}TiO_3$ but indicate a more rapid quenching of the $Ti^{3+}$ moment with doping level. For example the moment in the La system is already ~ 0.1 $\mu_B$ at $x$ ~ 0.05 while for the Nd materials the moment is still near 0.4 $\mu_B$ for the same $x$ (Figure 14). The greater resistance to quenching of the ordered $Ti^{3+}$ moment upon hole doping in the $Nd_{1-x}TiO_3$ system relative to that for $La_{1-x}TiO_3$ is likely due to increased levels of correlation, $U/W$, in the former.

Finally, the most unexpected result from this study is the observation of an extensive AF short-range order regime for $0.074(9) \leq x < 0.098(10)$ as evidenced by neutron diffraction (Figure 15) and the field dependence of the d.c. susceptibility (Figures 8b, c). This AF SRO region had not been found in previous studies on $Nd_{1-x}TiO_3$, the related $Nd_{1-x}Ca_xTiO_3$ system nor the $LaTiO_{3+\delta}$ system just described [4, 22, 24, 40]. The exact nature of this SR regime has not yet been characterized in detail. Note that there is now a closer parallel for the hole doped titanates with the hole doped cuprates in which the AF LRO region is separated by a spin-disordered region before the metallic state is reached as hole doping progresses. We have shown that the Anderson-Mott transition for $Nd_{1-x}TiO_3$ occurs over the same range of doping levels for which the AF SRO is found [26].

**Acknowledgements**

We wish to thank J.D. Garrett, H. Dabkowska and A. Dabkowksi for their assistance with single crystal preparations. Also, we would like to acknowledge P.C. Canfield at Iowa State University, for helpful discussions. Funding from Natural Science & Engineering Research Council of Canada in the form of a pre-doctoral scholarship for A.S. and a Discovery Grant for J.E.G. is also gratefully acknowledged.

**Table 1:** The NAA, TGA and cell volume data for a range of Nd$_{1-x}$TiO$_3$ samples.

| a | b | c | d | e | f | g |
|---|---|---|---|---|---|---|
| Target Nd/Ti ratio | Nd/Ti value by NAA | Cell volume (Å$^3$) | Theoretical % weight gain | Observed TGA % weight gain | Nd value modified by TGA | $x$ = 1 - column 'b' |
| 1 | 0.990(6) | 243.01(4) | 3.20 | 3.25(16) | 0.991(7) | 0.010(6) |
| 1 *(crystal)* | 0.981(6) | 243.05(5) | 3.10 | 3.22(16) | 0.982(7) | 0.019(6) |
| 0.98 | 0.943(11) | 240.90(4) | 2.75 | 2.87(14) | 0.945(11) | 0.057(11) |
| 0.97 | 0.917(1) | 239.90(4) | 2.50 | 2.90(15) | 0.923(3) | 0.083(1) |
| 0.965 | 0.926(9) | 240.07(4) | 2.59 | 2.90(15) | 0.931(10) | 0.074(9) |
| 0.96 | 0.911(1) | 239.45(5) | 2.45 | 2.72(14) | 0.916(2) | 0.089(1) |
| 0.96 | 0.921(2) | 239.87(4) | 2.55 | - | - | 0.079(2) |
| 0.95 | 0.905(8) | 239.10(4) | 2.39 | 2.61(13) | 0.908(9) | 0.095(8) |
| 0.93 | 0.888(4) | 238.91(3) | 2.22 | 2.63(13) | 0.894(4) | 0.112(4) |
| 0.92 | 0.875(4) | 237.71(5) | 2.10 | 2.29(11) | 0.878(4) | 0.125(4) |
| 0.90 | 0.874(9) | 236.91(5) | 2.10 | 2.22(11) | 0.876(9) | 0.126(9) |
| 0.85 *(crystal)* | 0.850(1) | 235.50(5) | 1.87 | 2.05(10) | 0.853(2) | 0.150(1) |

**Table 2:** Conditions for neutron data collection and profile refinements for structural determination from powder neutron data, at room temperature, for various vacancy-doped samples in $Nd_{1-x}TiO_3$.

| $x$ | 0.019(6) | 0.057(11) | 0.071(10) | 0.074(9) | 0.079(2) | 0.089(1) | 0.098(10) |
|---|---|---|---|---|---|---|---|
| Neutron λ (Å) | 1.330860 | 1.329170 | 1.329170 | 1.330350 | 1.330350 | 1.330048 | 1.330002 |
| Unit cell dimensions (Å) | 5.6570 (3) 7.7981 (5) 5.5272 (3) | 5.6120 (3) 7.8073 (5) 5.5165 (3) | 5.59996 (1) 7.80536 (1) 5.51103 (1) | 5.5955 (3) 7.8145 (5) 5.5136 (3) | 5.5894 (3) 7.8144 (5) 5.5112 (3) | 5.58259 (1) 7.80536 (2) 5.50352 (1) | 5.5725 (3) 7.8117 (5) 5.5024 (3) |
| Volume (Å³) | 243.826 (1) | 241.703 (1) | 240.8855 (2) | 241.088 (1) | 240.717 (1) | 239.8111 (2) | 239.523 (1) |
| Formula weight (g/mol) | 237.305 | 231.896 | 229.877 | 229.416 | 228.766 | 227.338 | 225.983 |
| $\chi^2$ | 5.73 | 2.58 | 2.54 | 2.41 | 2.63 | 2.54 | 4.68 |
| $R_p$ | 3.04 | 3.06 | 3.07 | 2.16 | 2.19 | 3.40 | 3.16 |
| $R_{wp}$ | 4.12 | 3.99 | 3.98 | 3.07 | 3.14 | 4.39 | 4.10 |
| $R_{exp}$ | 1.72 | 2.48 | 2.49 | 1.98 | 1.93 | 2.76 | 1.90 |
| $R_{Bragg}$ | 6.43 | 6.08 | 6.64 | 6.94 | 5.56 | 8.48 | 7.00 |
| $R_f$ | 4.51 | 4.64 | 4.51 | 5.97 | 5.36 | 6.70 | 5.28 |

**Table 3:** Atomic positions from the profile refinement of neutron powder data collected for $x$ compositions in $Nd_{1-x}TiO_3$, at room temperature.

|     |        | $x$ |  |  |  |  |  |  |
|-----|--------|-----------|------------|------------|-----------|-----------|-----------|------------|
|     |        | 0.019(6)  | 0.057(11)  | 0.071(10)  | 0.074(9)  | 0.079(2)  | 0.089(1)  | 0.098(10)  |
| Nd  | $x$    | 0.0604(4) | 0.0535(4)  | 0.0527(5)  | 0.0515(5) | 0.0502(5) | 0.0485(5) | 0.0467(5)  |
|     | $y$    | 0.25      | 0.25       | 0.25       | 0.25      | 0.25      | 0.25      | 0.25       |
|     | $z$    | 0.9895(6) | 0.9907(9)  | 0.9891(7)  | 0.9900(8) | 0.9897(8) | 0.9907(8) | 0.9922(8)  |
|     | $B$ (Å²) | 0.40(4) | 0.37(4)    | 0.11(4)    | 0.72(5)   | 0.64(5)   | 0.14(4)   | 0.35(4)    |
| Ti  | $x$    | 0.5       | 0.5        | 0.5        | 0.5       | 0.5       | 0.5       | 0.5        |
|     | $y$    | 0         | 0          | 0          | 0         | 0         | 0         | 0          |
|     | $z$    | 0         | 0          | 0          | 0         | 0         | 0         | 0          |
|     | $B$ (Å²) | 0.47(9) | 0.28(8)    | 0.12(8)    | 0.32(8)   | 0.36(8)   | 0.20(9)   | 0.28(8)    |
| O1  | $x$    | 0.4785(6) | 0.4810(6)  | 0.4818(6)  | 0.4817(7) | 0.4821(7) | 0.4826(7) | 0.4829(6)  |
|     | $y$    | 0.25      | 0.25       | 0.25       | 0.25      | 0.25      | 0.25      | 0.25       |
|     | $z$    | 0.0943(7) | 0.0877(7)  | 0.0877(8)  | 0.0872(8) | 0.0852(8) | 0.0818(8) | 0.0786(8)  |
|     | $B$ (Å²) | 0.63(7) | 0.73(6)    | 0.54(7)    | 0.92(7)   | 0.96(7)   | 0.67(7)   | 0.74(6)    |
| O2  | $x$    | 0.3014(5) | 0.2989(5)  | 0.2981(5)  | 0.2969(5) | 0.2964(5) | 0.2963(5) | 0.2948(4)  |
|     | $y$    | 0.0483(3) | 0.0463(3)  | 0.0453(3)  | 0.0437(4) | 0.0437(4) | 0.0442(4) | 0.0427(4)  |
|     | $z$    | 0.6992(5) | 0.7029(5)  | 0.7030(5)  | 0.7042(5) | 0.7040(5) | 0.7041(5) | 0.7052(5)  |
|     | $B$ (Å²) | 0.64(5) | 0.52(5)    | 0.42(5)    | 0.88(5)   | 0.84(5)   | 0.45(5)   | 0.58(5)    |

**Table 4**: The heat capacity data summary for two $Nd_{1-x}TiO_3$ single crystal samples.

|  | $T_N$ (K) | Region integrated (K) | Entropy S J/(mol.K) |
|---|---|---|---|
| $Nd_{0.981(6)}TiO_3$ | 88.3 | 60 – 97 | 1.835 |
| $Nd_{0.936(10)}TiO_3$ | 61.9 | 42 - 80 | 1.169 |

**Table 5:** Summary of the ordering temperature ($T_{order}$) obtained from the disappearance of remanent moments analyses, M(T). The $T_N$ are found from heat capacity C(T) data and estimated from $d(\chi T)/dT$ analyses. The divergence temperature between the zfc/fc samples, $T_D$, is also summarized for various $Nd_{1-x}TiO_3$ samples.

| x | NTO formula | C(T) $T_N$ (K) | $d(\chi T)/dT$ ~$T_N$ ± 1 (K) | zfc/fc $\chi(T)$ $T_D$ ± 5 (K) | M(T) $T_{order}$ ± 2 (K) | calculated [$Ti^{3+}$] |
|---|---|---|---|---|---|---|
| 0.010(6) | $Nd_{0.990(6)}TiO_3$ | - | - | 98 | 102 | 0.970(18) |
| 0.019(6) | $Nd_{0.981(6)}TiO_3$ | 88.2 | 86 | 95 | 87 | 0.943(18) |
| 0.057(11) | $Nd_{0.943(11)}TiO_3$ | - | 70 | 90 | 80 | 0.829(33) |
| 0.064(10) | $Nd_{0.936(10)}TiO_3$ | 61.9 | 61 | 70 | 66 | 0.808(30) |
| 0.071(10) | $Nd_{0.929(10)}TiO_3$ | - | 60 | 80 | 75 | 0.787(30) |
| 0.074(9) | $Nd_{0.926(9)}TiO_3$ | - | 41, 46 | 52, 58 * | 58 | 0.778(27) |
| 0.079(2) | $Nd_{0.921(2)}TiO_3$ | - | - | 38 | 35 | 0.763(6) |
| 0.080(10) | $Nd_{0.920(10)}TiO_3$ | - | - | 42, 70 * | - | 0.760(30) |

*: These samples give different ordering temperatures at various applied fields.

**Table 6:** Refinement results for the combined chemical and magnetic *Pnma* structures of $Nd_{1-x}TiO_3$ samples at ~ 4 K. The magnetic moments listed here are the averaged values per Ti (and Nd) site, with $G_x$ on $Ti^{3+}$ and $C_y$ on $Nd^{3+}$.

| $x$ | 0.019(6) | 0.034(10) | 0.057(11) | 0.071(10) | 0.074(9) |
|---|---|---|---|---|---|
| Sample | $Nd_{0.981(6)}TiO_3$ | $Nd_{0.966(10)}TiO_3$ | $Nd_{0.943(11)}TiO_3$ | $Nd_{0.929(10)}TiO_3$ | $Nd_{0.926(9)}TiO_3$ |
| Neutron $\lambda$ (Å) | 2.37298 | 2.369640 | 2.369570 | 2.378607 | 2.371886 |
| *Magnetic moment* ($\mu_B$) $Nd^{3+}$ | 0.829 (23) | 0.741 (25) | 0.761 (18) | 0.736 (20) | 0.614 (27) |
| $Ti^{3+}$ | 0.453 (75) | 0.374 (44) | 0.245 (36) | 0.247 (39) | 0.058 (109) |
| $\chi^2$ | 7.71 | 2.45 | 3.77 | 11.9 | 6.78 |
| $R_p$ | 3.57 | 2.79 | 2.79 | 3.13 | 2.82 |
| $R_{wp}$ | 4.80 | 3.55 | 3.66 | 4.11 | 3.92 |
| $R_{Bragg}$ | 4.38 | 2.14 | 2.45 | 2.30 | 2.99 |
| $R_f$ | 4.66 | 2.70 | 2.70 | 2.40 | 2.91 |
| $R_{mag}$ | 15.7 | 11.2 | 11.3 | 10.4 | 20.4 |

$$R_{Bragg} = 100 \sum_j |I_{jo} - I_{jc}| / \sum_j I_{jo}, \quad R_p = 100 \sum_i |y_{jo} - y_{jc}| / \sum_i y_{jo}, \quad R_{wp} = 100 \left[ \sum_i w_i (y_{io} - y_{ic})^2 / \sum_i w_i y_{io}^2 \right]^{1/2},$$

$$R_{exp} = 100 \left[ (N - P + C) / \sum_i w_j y_{io}^2 \right]^{1/2}, \quad \chi^2 = [R_{wp} / R_{exp}]^2$$

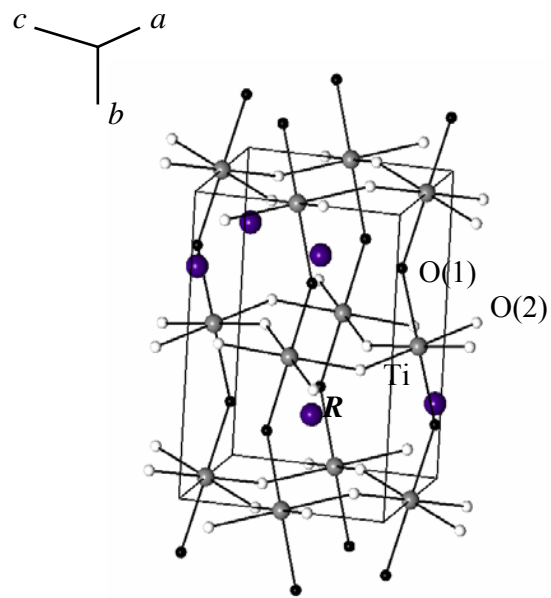

**Figure 1**: The structure of GdFeO$_3$-distorted $R$TiO$_3$ perovskite in *Pnma* symmetry.

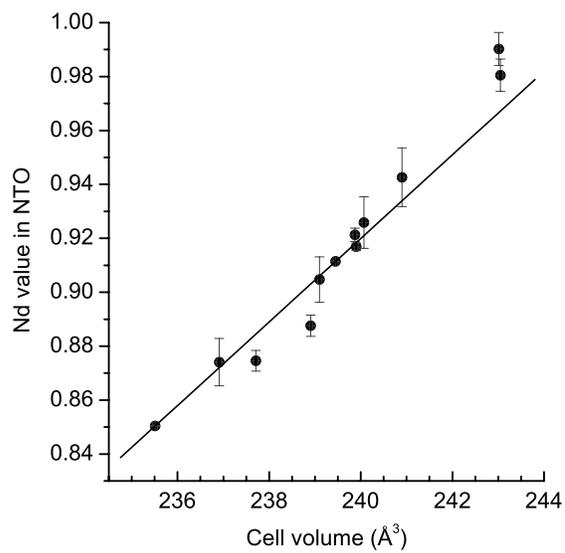

**Figure 2:** The plot of the neodymium values versus cell volume for various Nd$_{1-x}$TiO$_3$ compositions.

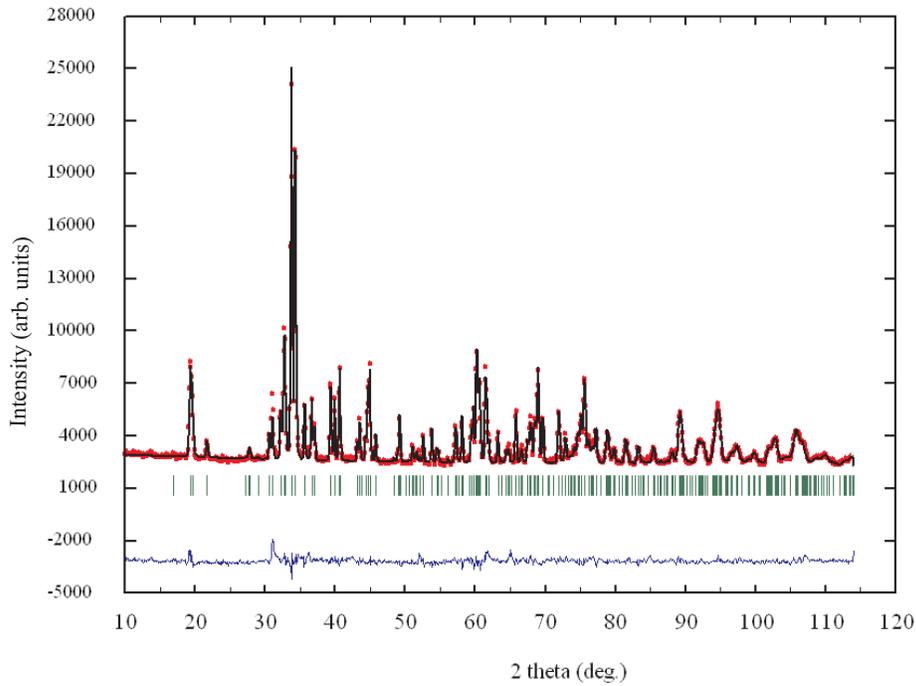

**Figure 3**: Refined neutron powder diffraction profiles for $Nd_{0.981(6)}TiO_3$, with $x = 0.019(6)$; see refinement results in Table 2.

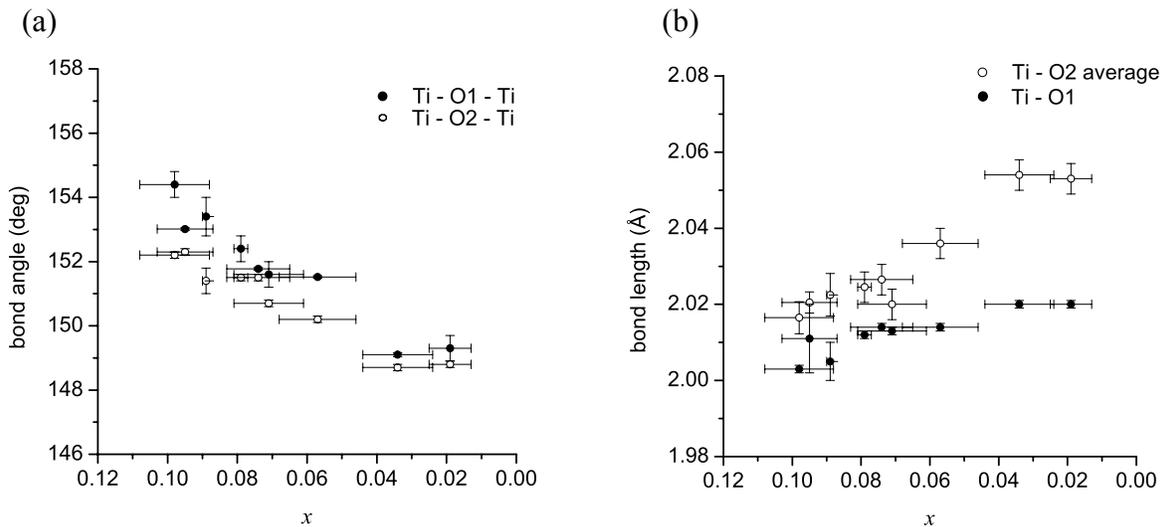

**Figure 4:** The (a) Ti – O – Ti bond angles and (b) Ti – O bond distances as a function of neodymium vacancies, $x$, in $Nd_{1-x}TiO_3$ solid solution.

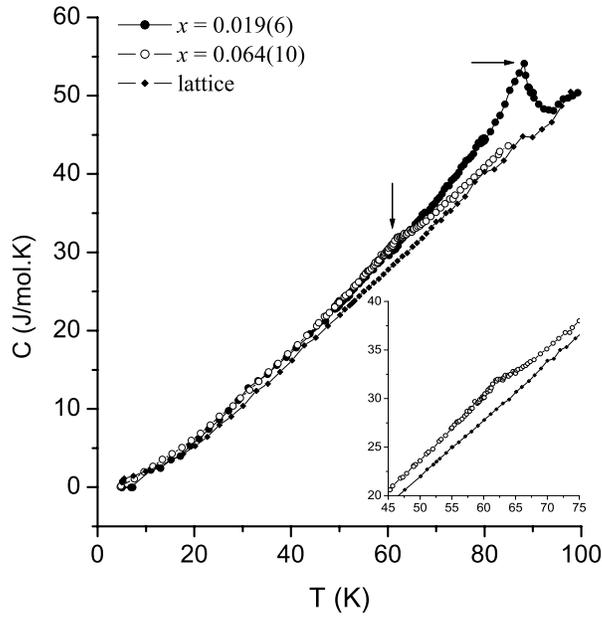

**Figure 5**: The heat capacity vs temperature for $Nd_{0.981(6)}TiO_3$ with $x = 0.019(6)$, $Nd_{0.936(10)}TiO_3$ with $x = 0.064(10)$, and $Nd_{0.89}TiO_3$ sample with $x \cong 0.11$.

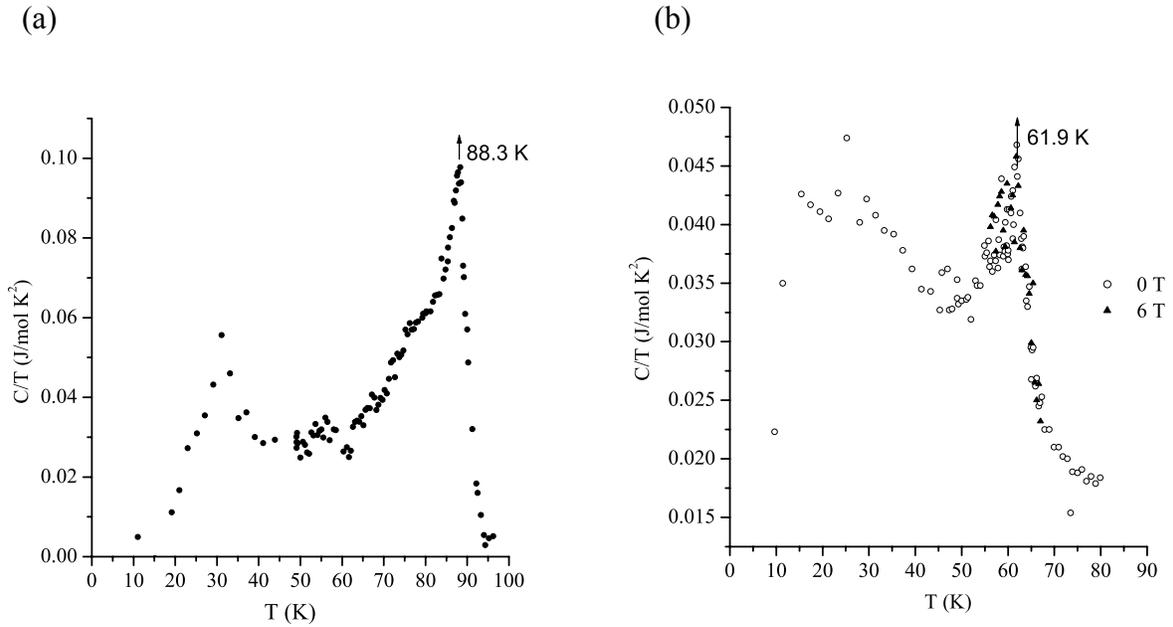

**Figure 6**: The C/T vs temperature result for the magnetic contribution of heat capacity of (a) $Nd_{0.981(6)}TiO_3$, with $x = 0.019(6)$ and (b) $Nd_{0.936(10)}TiO_3$, with $x = 0.064(10)$.

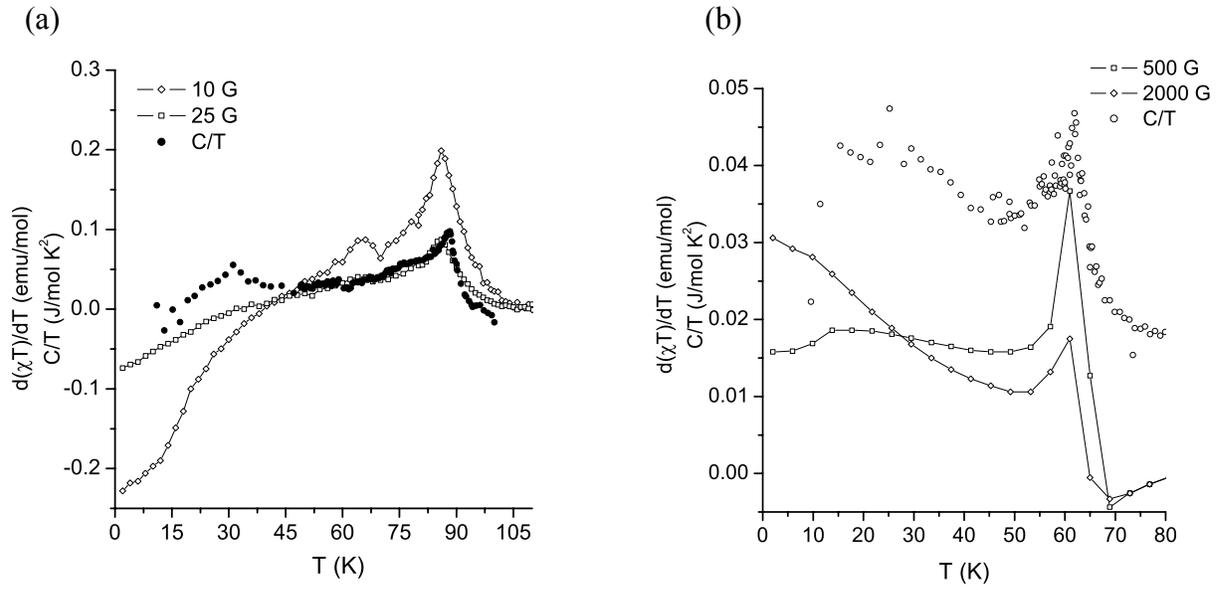

**Figure 7**: The magnetic contribution of heat capacity (C/T) and magnetic susceptibility $d(\chi T)/dT$ in (a) $x = 0.019(6)$ and (b) $x = 0.064(10)$ samples.

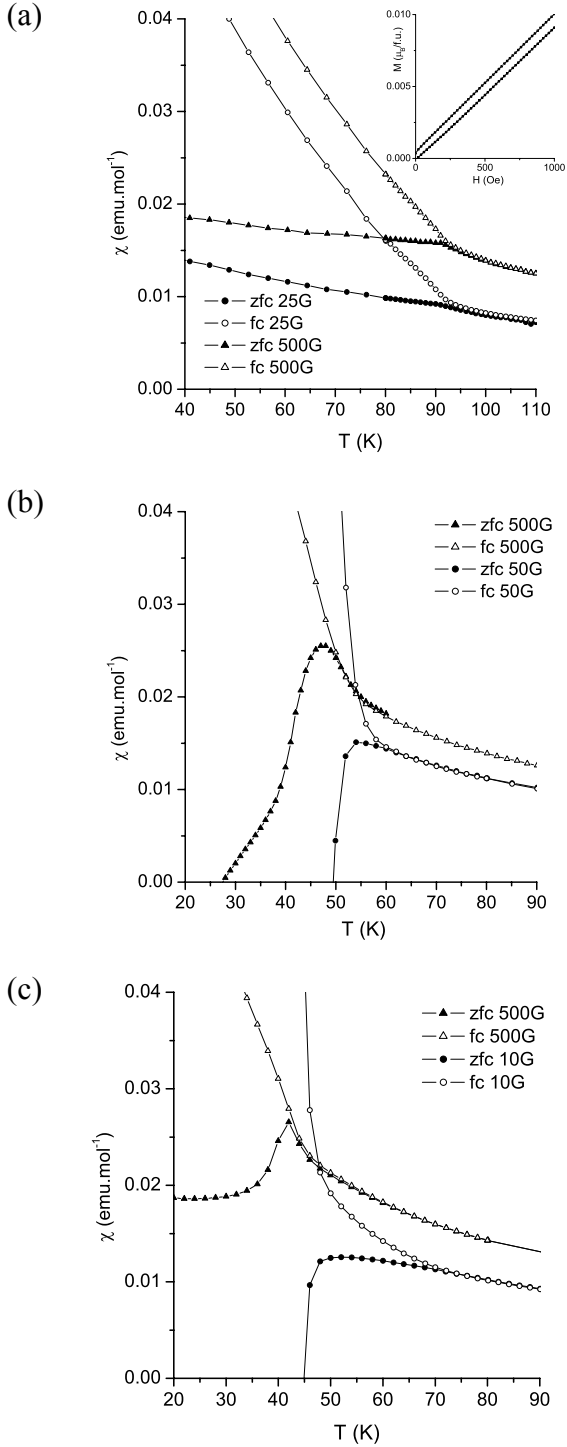

**Figure 8**: (a) The susceptibility data for $Nd_{0.981(6)}TiO_3$ with $x = 0.019(6)$. The region near the divergence is shown to have no dependence on the applied field. The inset is the magnetization vs applied field at 2 K. (b) The susceptibility for $Nd_{0.926(9)}TiO_3$, $x = 0.074(9)$ (c) and $Nd_{0.920(10)}TiO_3$, $x = 0.080(10)$; the region near the divergences is shown, having dependence on the applied field.

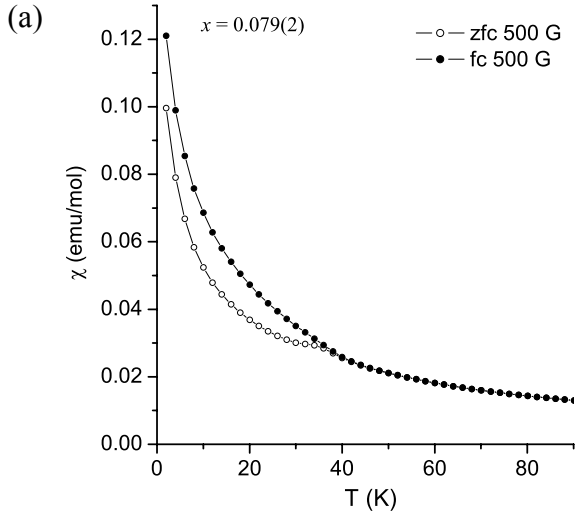
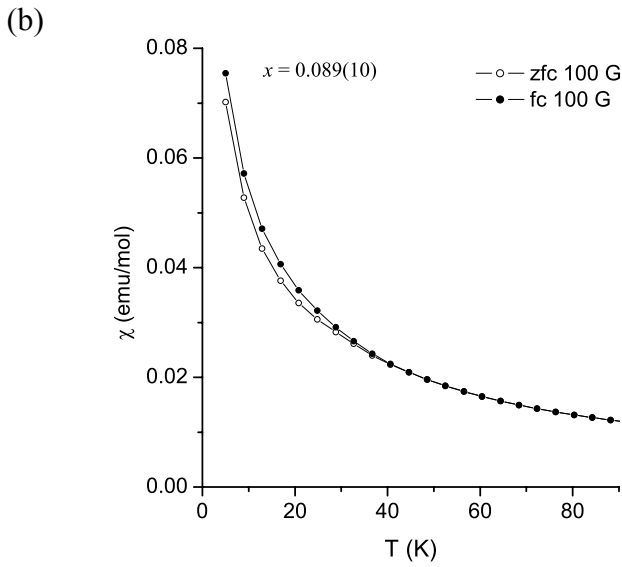
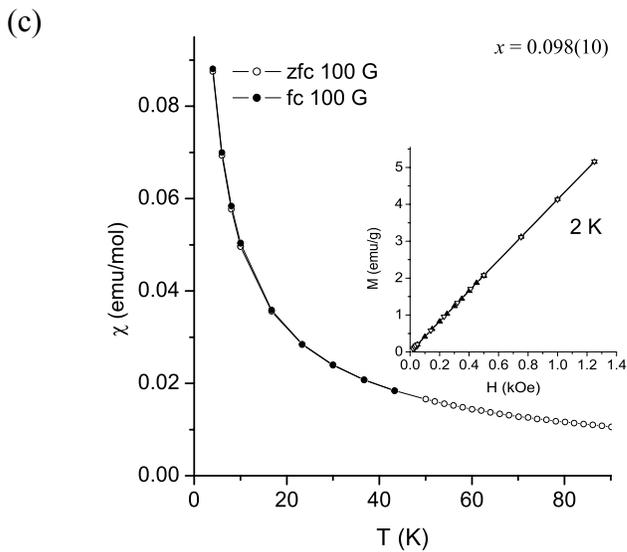

**Figure 9**: The molar magnetic susceptibility data for (a) $Nd_{0.921(2)}TiO_3$ with $x = 0.079(2)$, (b) $Nd_{0.916(2)}TiO_3$ with $x = 0.089(10)$, and (c) $Nd_{0.902(10)}TiO_3$ with $x = 0.098(10)$. The inset of (c) is the magnetization data at 2 K.

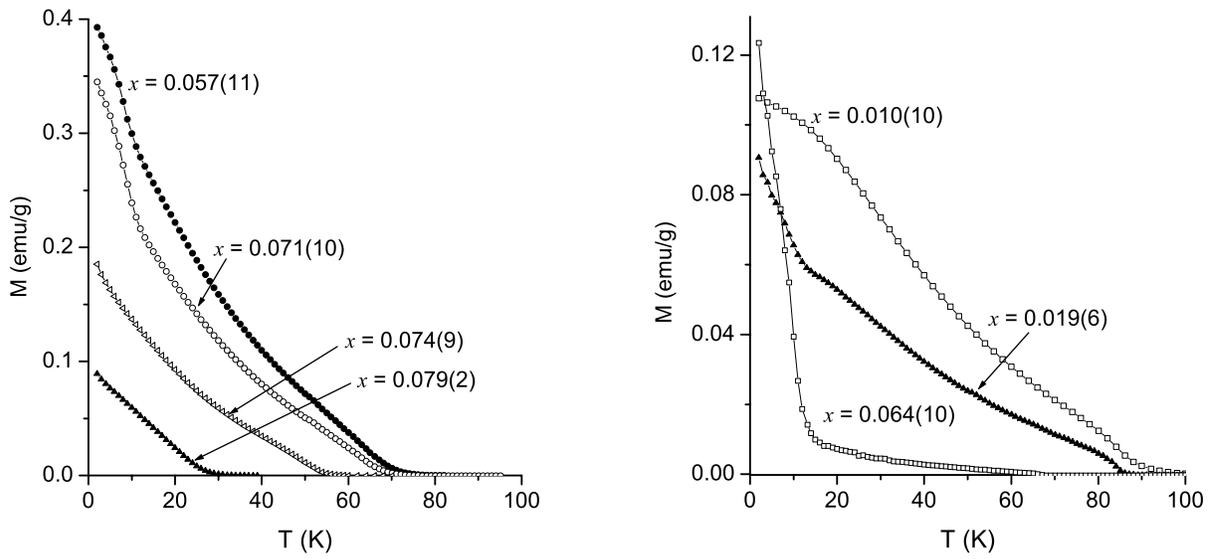

**Figure 10**: The plot of magnetization versus temperature in the composition range $0.010(10) \leq x \leq 0.125(4)$ for $Nd_{1-x}TiO_3$; $x$ values are shown on the graph.

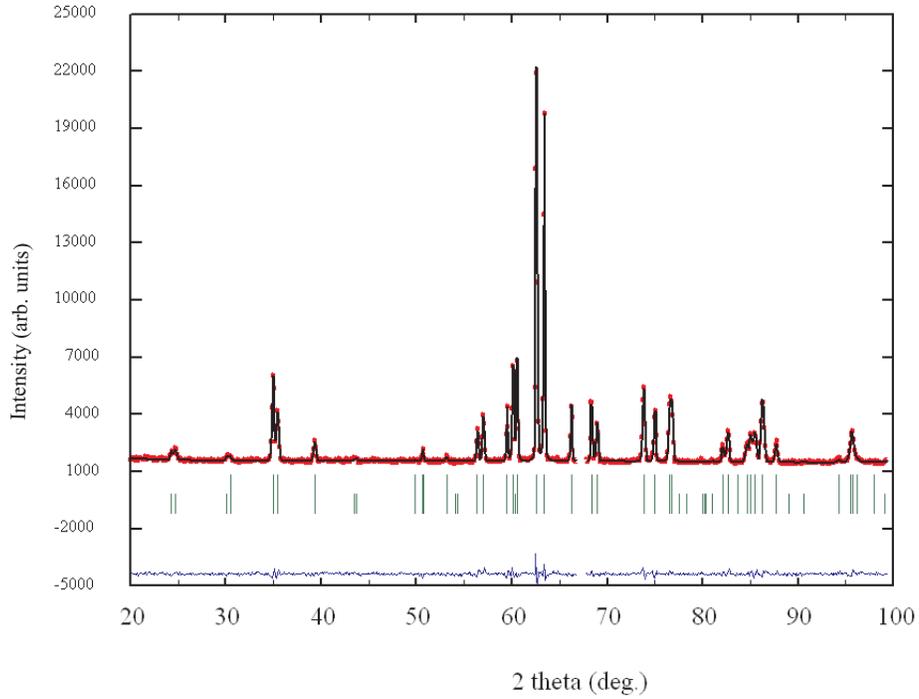

**Figure 11**: Rietveld refinement of $Nd_{0.981(6)}TiO_3$ at 4 K using $\lambda = 2.37$ Å neutrons of the crystal and magnetic structures with $G_x$ on $Ti^{3+}$ and $C_y$ on $Nd^{3+}$. The strongest magnetic reflections are marked; see refinement results in Table 6.

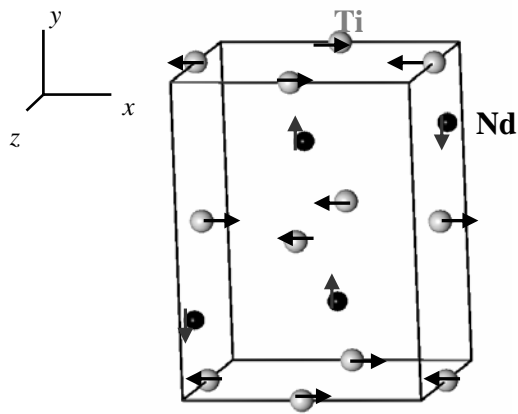

**Figure 12:** The arrangement of spins in the $G_x$-type ($Ti^{3+}$) and $C_y$-type ($Nd^{3+}$) configurations found for the magnetically ordered $Nd_{1-x}TiO_3$ *Pnma* phases at 4K.

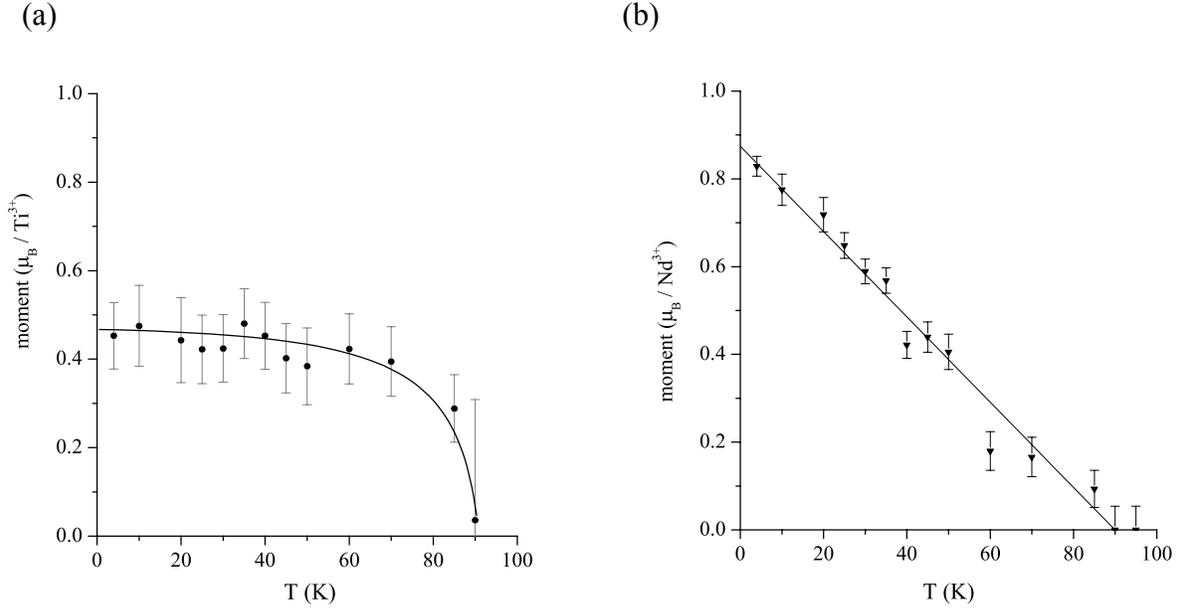

**Figure 13:** Magnetic moment vs temperature for $Nd_{0.981(6)}TiO_3$ with $x = 0.019(6)$. The refinement is in *Pnma* setting and $G_x$ configuration on $Ti^{3+}$ and $C_y$ configuration on $Nd^{3+}$.

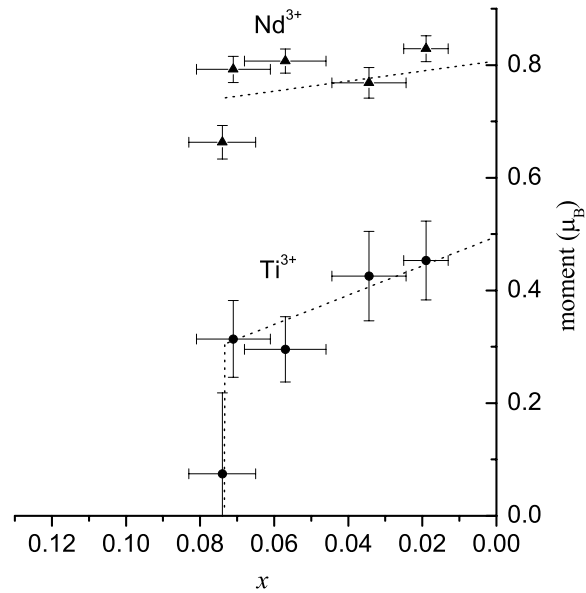

**Figure 14:** The refined magnetic moments per $Ti^{3+}$ ($G_x$) and $Nd^{3+}$ ($C_y$) up to $x = 0.074(9)$ sample composition in $Nd_{1-x}TiO_3$ *Pnma*. The concentrations of $Ti^{3+}$ and $Nd^{3+}$ have been inferred from the analytically determined formula.

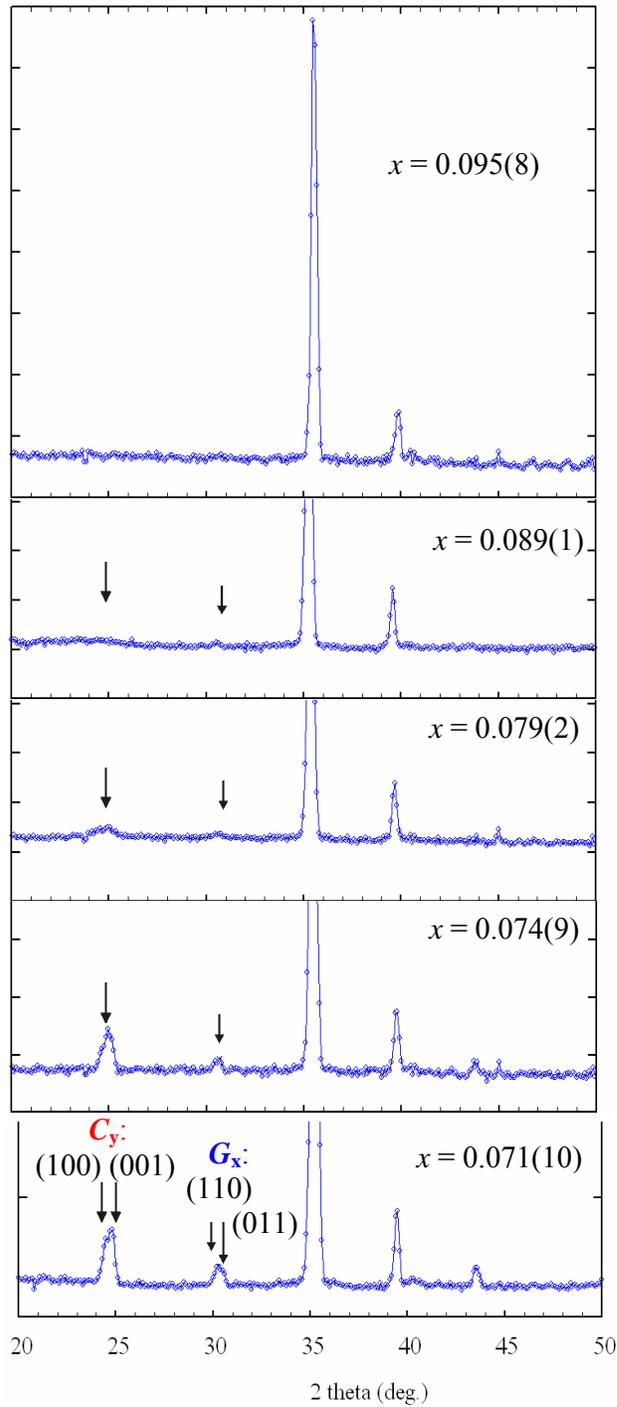

**Figure 15:** The sharp Bragg reflections in $x = 0.071(10)$, indicated by arrows, give way to broadened peaks indicative of short-range ordering in $x = 0.074(10)$ to $0.089(1)$ $Nd_{1-x}TiO_3$ samples. No magnetic features are evident for $x = 0.095(8)$.

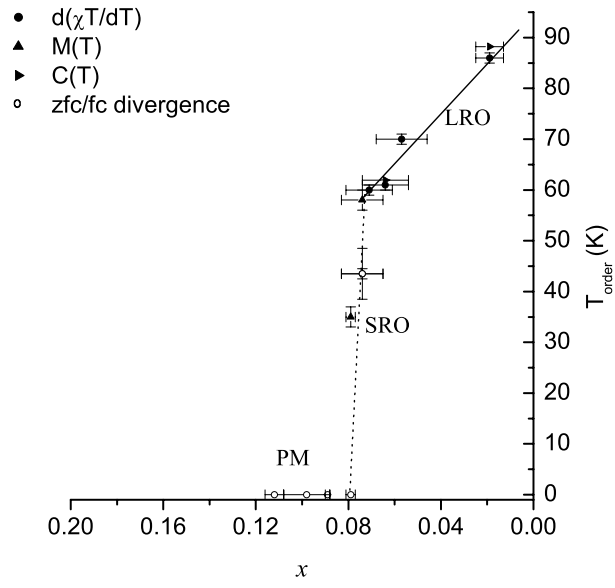

**Figure 16**: The ordering temperature versus $x$ for samples in the $Nd_{1-x}TiO_3$ system. Regions of AF LRO, AF SRO and paramagnetic (PM) behavior are indicated.